\documentclass[pra,twocolumn,showpacs,aps,amssymb]{revtex4}
\usepackage{amssymb}

\usepackage{graphicx}
\usepackage{amsmath}
\usepackage{times}

\begin{document}

\title{Bragg Spectroscopy of ultracold atoms loaded in an optical lattice}
\author{$^{(1,2)}$Ana Maria Rey ,$^3$P. Blair Blakie, $^{(1,2)}$Guido Pupillo , $^1$%
Carl J. Williams and $^1$Charles W. Clark .}
\date{\today}
\affiliation{$^{1}$ Physics Laboratory, National Institute of
Standards and Technology, Gaithersburg, MD 20899- 8410, USA.}
\affiliation{$^{2}$ Department of Physics, University of Maryland,
College Park, USA.} \affiliation{$^{3}$ Department of Physics,
University of Otago, P.O Box 56 Dunedin, New Zealand.}

\begin{abstract}
We study Bragg spectroscopy of ultra-cold atoms in one-dimensional
optical lattices as a method for probing the excitation spectrum in
the Mott insulator phase, in particular the one particle-hole
excitation band. Within the framework of perturbation theory  we
obtain an  analytical expression for the dynamic structure factor
$S(q,\omega )$ and use it to calculate the imparted energy which has
shown to be a relevant observable  in recent experiments. We test
the accuracy of our approximations by comparing them with
numerically exact solutions of the Bose-Hubbard model in restricted
cases and establish  the limits of validity of our linear response
analysis. Finally we show that when the system is deep in the Mott
insulator regime, its response to the Bragg perturbation is
temperature dependent. We suggest that this dependence might be used
as a tool to probe temperatures of order of the Mott gap.
\end{abstract}

\maketitle

\section{Introduction}

Recently, there has been a lot of experimental progress studying
cold atoms confined in optical lattices. The  defect free nature
of the lattice potential,  the long coherence times of the
constituent atoms,  and the high experimental control of the
lattice parameters \cite{Denschlag2002a,Peil2003a} make this a unique system for precisely studying
many-body physics. In particular, the experimental observation of the superfluid to Mott insulator
quantum phase transition \cite{Greiner2002a} has stimulated much interest in this area of research.

Perhaps one of the most important potential applications  of the
Mott insulator transition  is to use it as a means to initialize a
quantum computer
register\cite{Deutsch1998a,Calarco2000a,Jaksch2003a,BHdynamics}.
Particulary in this case, it is  important to have tools for
thoroughly characterizing the experimentally obtained Mott
insulator states. The usual procedure for entering the Mott
insulator regime is to begin with a magnetically trapped BEC (with
almost all the atoms in the condensate), slowly load it into an
optical lattice by increasing the lattice depth. One key  piece of
evidence for the quantum phase transition is the loss of global
phase coherence of the matter wavefunction when the lattice depth
increases beyond a critical value \cite{Greiner2002a}. However,
the loss of coherence could arise from many sources, such as the
decoherence induced by quantum or thermal depletion of the
condensate during the loading process \cite{Rey2003a} and
therefore it is not a sufficient signature that the system is in
the Mott insulator state.
 For this reason, in the  experiments by Greiner \emph{et
al.}\cite{Greiner2002a}, complementary evidence for the Mott
insulator transition was provided by applying a potential gradient
to the lattice to show the presence of a gap in the excitation
spectrum. In this paper we show that Bragg spectroscopy, done by
applying  additional laser beams independent from the lattice beams,
is an experimental technique with the potential to thoroughly
characterize the Mott phase. In addition to determining the energy
gap, we show that it  provides detailed information about the
excitation spectrum, information unavailable using other techniques.
Moreover, in contrast to applying a potential across the lattice
Bragg spectroscopy is not susceptible to effects like Bloch
oscillations and Zener tunneling. Furthermore, we show that
different from the superfluid regime, the system's response to Bragg
perturbation in the Mott regime is sensitive to finite temperature.
This property might be used as a tool to probe temperatures of order
of the Mott gap.

Our analysis is based on a perturbative treatment, which we show
to be applicable in the strong Mott regime that has been reached in
 current experiments \cite{Paredes2004}.
Although our approach is applicable only in the range of validity
of first order perturbation theory, it has the advantage of
properly including one-particle-hole correlations.  Such
correlations  have dominant influence on the spectrum of the
system and are not accounted for in mean-field treatments
\cite{Stoof2004}.

The organization of this paper is the following. In
Sec.~\ref{Form} we introduce the basic formalism that describes
Bragg spectroscopy in an optical lattice  and use a linear
response approach to calculate the imparted energy to the system.
In Sec.~ \ref{Zero} we derive the zero temperature response  to
Bragg spectroscopy of a translationally invariant lattice deep in
the Mott regime and in Sec.~\ref{validity} we  discuss the
conditions  required for our linear response analysis to be valid.
In Sec.~\ref{fin} we extend the zero temperature analysis to
finite temperature and finally in Sec.~\ref{con} we conclude.

\section{Formalism}
 \label{Form}

The typical Bragg spectroscopic procedure is to gently scatter
atoms in an ultra-cold gas system with a moving potential of the
form $V_{0}\cos (qx-\omega t)$.  This type of experiment was first
demonstrated  by the
 NIST and MIT groups
\cite{Stenger1999a,StamperKurn1999a}. In contrast to earlier
experiments that used momentum as the response observable, here we
choose to examine the imparted energy. In trapped systems this
allows long excitation duration that facilitates more precise
spectral resolution. Energy-spectroscopy is not as well developed
as momentum spectroscopy but recent experiments have demonstrated
the use of this technique \cite{Stoferle2004a}. The Bragg
potential is formed by the ac-Stark shift arising from a pair
 of interfering light fields (e.g. see \cite{Blakie2002a}).
In this paper we will always assume that the Bragg potential is
\textit{generated independently} of and is much weaker than the
lattice
 potential.  We therefore treat the scattering process with linear response theory.
Using an independent set of beams to generate the Bragg potential
also provides considerable flexibility in the range
 of $q$ and $\omega$ values that can be obtained.

In this work we consider one-dimensional bosons loaded in an
optical lattice. Effective 1D systems have been realized in recent
experiments (for example refs. \cite{Stoferle2004a,Paredes2004})
by loading a Bose-Einstein condensate into a three-dimensional
optical lattice, which is very deep in two directions. The
dynamics is then restricted to the third, or axial, direction
only. In this work we study the response of the system to Bragg
perturbation in the axial direction, assuming that the dynamics in
the transverse directions is frozen. We consider a one-dimensional
optical lattice which is sufficiently deep that the tight binding
approximation is valid and assume that we can restrict the
dynamics of the atoms to the lowest vibrational band. This applies
when changing the lattice potential does not induce band
excitations. This condition is fulfilled when the frequency,
$\omega$, of the Bragg perturbation is less than the gap between
the first and second bands, and when the momentum transfer $q$ is
contained within the first Brillioun zone. A detailed analysis of
the validity of this first-band approximation to study Bragg
scattering of a dilute weakly-interacting gas in an optical
lattice is found in \cite{Menotti2003a} where the authors used a
mean field approach combined with Bogoliubov analysis. In the
single-band approximation and in absence of external potentials,
the system is described by the Bose-Hubbard Hamiltonian (BHH)
\cite{Jaksch1998a}
\begin{eqnarray}
\hat{H} &=&-\!J\!\sum_{\langle n,l\rangle} \hat{a}_{n}^{\dagger
}\hat{a}_{l} +\frac{U}{2} \sum_{n}\hat{a}_{n}^{\dagger
}\hat{a}_{n}^{\dagger }\hat{a}_{n}\hat{a}_{n}. \label{EQNBHH}
\end{eqnarray}
Here $\hat{a}_{n}$ is the annihilation operator at site $n$ which
obeys the canonical commutation relations for bosons, $J$ is the
hopping matrix element between nearest neighbors, and $U$ is the
on-site repulsion energy. The sum $\langle n,l\rangle$ is taken over
nearest
 neighbors. We use $N$ for the total number of atoms and $M$ for the
 total number of wells. In the tight binding approximation the
Hamiltonian describing the Bragg perturbation reads
\begin{equation}
\hat{H}_{B}=\frac{1}{2}V_{0}(\hat{\rho}_{q}^{\dagger }e^{-i\omega t}+\hat{%
\rho}_{q}e^{i\omega t}),
\end{equation}
where the density fluctuation operator $\hat{\rho}_{q}^{\dagger }$
is defined as $\hat{\rho}_{q}^{\dagger
}=\sum_{n,m=0}^{M-1}I_{q}^{n-m}\hat{a}_{m}^{\dagger }\hat{a}
_{n}e^{iqmd}$, where $ I_{q}^{n}\!=\!\int dx e^{iqx}\phi
_{0}^{\ast }(x) \phi _{0}(x-dn)$ is a geometrical factor that
involves integration over Wannier functions $\phi_{0}(x)$, and $d$
is the lattice constant. For deep lattices $I_{q}^{n}\propto
\delta_ {n,0}$ \cite{Menotti2003a}.

To analyze the Bragg spectrum of the system we study the energy
transfer, which can be measured by time-of-flight
techniques\cite{Stoferle2004a}. Under linear response theory, the
energy transfer is related to the so called {\it dynamic structure
factor}, $S(q,\omega)$, which is given by
 \begin{eqnarray}
S(q,\omega )\, &=&\frac{1}{\mathcal{Z}}\sum_{ij}e^{-\beta
E_{i}}f_q(\omega_{ij}), \label{dyn}
\end{eqnarray}
where $f_q(\omega_{ij})\equiv \left| \left\langle i\right|
\hat{\rho}_{q}\left| j\right\rangle \right| ^{2}\delta (\omega
-\omega _{ij})$, $\left| i\right\rangle $ and $E_{i}$ are
eigenstates and eigenenergies of the unperturbed Hamiltonian
(\ref{EQNBHH}), $e^{-\beta E_{i}}$ is the usual Boltzmann factor
with $\beta =1/k_B T$ where $ k_B $ is Boltzmann's constant and
$T$ the temperature, $\mathcal{Z}$ is the canonical partition
function, and $\hbar\omega_{ij}=E_{j}-E_{i}$. Because of the
factor $f_q(\omega_{ij})$, the system's response shows peaks
whenever the frequency of the Bragg perturbation matches the
energy difference between two eigenstates of the BHH. The peak
height is proportional to the the transition probability  between
the two eigenstates, $\left| \left\langle i\right|
\hat{\rho}_{q}\left| j\right\rangle \right|^2$.

The total energy transfer after applying the Bragg perturbation
can be shown  to be given by \cite{Brunello2001a}:
\begin{eqnarray}
\delta E &=&\frac{V_{0}^2}{2\hbar}\int_0^ {T_p}
dt\int_{-\infty}^{\infty} d\omega'\omega' \chi(q,\omega')
\frac{\sin((\omega-\omega')t)}{(\omega-\omega')}, \label{impene}
\end{eqnarray}
where $T_p$ is the duration of the perturbation and
$\chi(q,\omega)=S(q,\omega)-S(-q,-\omega)$. Here we derive
analytic expressions for the dynamic structure factor assuming we
are deep in the Mott insulator regime, where treating the hopping
term in the Hamiltonian as a perturbation is justified.
\section{Zero temperature response}
 \label{Zero}
 In this
work  we assume a commensurately filled lattice  with no external
confinement, filling factor $N/M=g$ and periodic boundary
conditions. The unperturbed Hamiltonian includes only the on-site
interaction term, which is diagonal in a number Fock-state basis. To
zeroth order the ground state $| \Phi _{0}^{{(0)}}\rangle $ is the
Fock state with $g$ atoms in every lattice site. The lowest lying
excitations correspond to the one-particle-hole (1-ph) states
$\left| \Psi _{mn}\right\rangle$ with $g+1$ particles at site $m$,
$g-1$ particles at site $n$, and exactly $g$ particles in every
other site. There are $M(M-1)$ 1-ph excitations and, because of the
translational symmetry, they are degenerate at zeroth order with
excitation energy $U$. To zeroth order the dynamic
 structure factor vanishes. At
first order the ground state wave-function is $| \Phi
_{0}^{(1)}\rangle = | \Phi _{0}^{(0)}\rangle  +J/U \sqrt{2M
g(g+1)} |S \rangle$, where $|S\rangle\equiv\sum_{n=1}^{M}(| \Psi
_{nn+1}\rangle +| \Psi _{nn-1}\rangle)/\sqrt{2M}$ is the
normalized translationally invariant state of adjacent
particle-hole excitations. In order for perturbation theory to be
valid, the parameter $J g \sqrt{M}/U$ has to be small
\cite{BHdynamics}. This could be  a significant restriction for
systems with a large number of filled sites but can be perfectly
realized in experiments such as Ref.\cite{Paredes2004} where the
system has  only 20 occupied sites in the central tube. To find
first order corrections to the $M(M-1)$ low lying excited states
we must diagonalize the kinetic energy Hamiltonian within the 1-ph
subspace. If we expand the eigenstates as a linear combinations of
1-ph excitations $ | \Phi _{i}^{(1)}\rangle =\sum_{n,m \neq
n}c_{nm}^{i}\left| \Psi _{nm}\right\rangle$
 the necessary and sufficient conditions that the
coefficients $c_{\small{nm}}^{i}$ have to fulfill  are
\begin{equation}
(g+1)(c_{\small{n+1m}}^{i}+c_{\small{n-1m}}^{i})+g(c_{\small{nm+1}}^{i}+c_{\small{nm-1}}^{i})=
\widetilde{E}_{i}c_{\small{nm}}^{i},  \label{eig}
\end{equation}
with  $E_{i}^{(1)}=U-J\widetilde{E}_{i}$. Besides Eq. (\ref{eig}),
the amplitudes $c_{nm}^{i} $ have to satisfy periodic boundary
conditions $c_{n+Mm}^{i} = c_{nm+M}^{i}= c_{nm}^{i}$ and the
constraint $c_{nn}^{i} = 0 $ (which prevents particle and hole
excitations occurring at the same site).
 Eq. (\ref{eig}) is analogous to the tight
binding Schr\"odinger equation of a two dimensional square lattice
in the $xy$-plane. The $x$ direction is associated with the position
of the extra particle  and the $y$  direction with the position of
the hole. The different weights $g+1$ and $g$ can be understood in
the 2D-lattice model as different effective masses in the two
directions and the constraint $c_{nn}^{i} = 0$ as a hard wall along
the $x=y$ line. The solutions are not straightforward
 due to the fact that the effective mass difference breaks
the lattice symmetry around the $x=y$ axis and makes the hard wall
constraint hard to fulfill. However, in the limiting case of high
filling factor, $g\gg 1$ the solutions of Eq. (\ref{eig})
(including the constraints) are
\begin{eqnarray}
E_{rR}^{(1)}&=&U-2J(2g+1)\cos \left( \frac{\pi r}{M}\right) \cos
\left( \frac{\pi R}{M}\right), \label{Spect}\\
 c_{nm}^{r,R\neq 0}
&=&\!\left\{
\begin{array}{c}
\frac{2}{M}\sin ( \frac{\pi r}{M}|n\!-\!m|) \sin ( \frac{%
\pi R^{\prime}}{M}(n\!+\!m)\!+\!\alpha_{\small{rR}}), \\
\frac{2}{M}\sin ( \frac{\pi r}{M}(n\!-\!m)) \sin ( \frac{\pi R^{\prime \prime}}{M%
}(n\!+\!m)\!+\!\beta_{\small{rR}}) ,
\end{array}
\right.     \\
c_{nm}^{r,0} &=&\!\left\{
\begin{array}{c}
\frac{\sqrt{2}}{M}\sin \left( \frac{\pi r}{M}|n\!-\!m|\right) ,\quad r\quad \rm{odd} \\
\frac{\sqrt{2}}{M}\sin \left( \frac{\pi r}{M}(n\!-\!m)\right)
,\quad r\quad \rm{even}\label{sol2}
\end{array}
\right.
\end{eqnarray}
Where we used $i=(r,R)$, with $r=1,\dots M-1$ and $R=0,\dots M-1$.
The notation $ R^{\prime}$ restricts the values of $R$ to the ones
where $R+r$ is an odd number and $ R^{\prime \prime}$ to the
values where $R+r$ is even. The constants $\alpha_{\small{rR}}=\pi
(r\!-\!R\!+\!1)/4$ and $\beta_{\small{rR}}=\pi
(r\!+\!R\!-\!1\!+\!M)/4$  guarantee the orthogonality of the
eigenmodes.

Using Eqs. (\ref{Spect}) and (\ref{sol2}) we get an expression for
the zero temperature dynamic structure factor given by
\begin{eqnarray}
&&S_0(q,\omega) =\frac{J^2}{U^{2}} g(g\!+\!1)
\sum_{r,R}\delta(\omega-E_{rR}^{(1)}/\hbar) \label{something} |
\sum_{m=1}^{M}e^{iqdm}H_m^{rR}
| ^{2} \notag \\
&&=\label{contr} 32\frac{J^2}{U^{2}} g(g\!+\!1)
 \sin^2( \frac{q d}{2}) {\sum_{r}}%
^{\prime} \sin ^{2}(\frac{ \pi r}{M})\delta
(\omega-\frac{E_{r\widetilde{q}}^{(1)}}{\hbar}).
\end{eqnarray}
Where $H_m^{rR}=
c_{mm+1}^{rR}\!+\!c_{mm-1}^{rR}\!-\!c_{m+1m}^{rR}\!-\!c_{m-1m}^{rR}$,
$qd=2\pi\widetilde{q}/M $, and $\widetilde{q}$ an integer between
$0$ and $M-1$. The prime in the sum imposes the constraint
$\widetilde{q}+r$ is even. It is important to emphasize that only
the states with $R=0$ have a dispersion relation which agrees to
first order in $J$ to the mean-field solution found in Ref.
\cite{Stoof2004}. However, for these states $H_m^{r0}=0$.

\begin{figure*}[tbh]
\begin{center}
\leavevmode {\includegraphics[width=6. in]{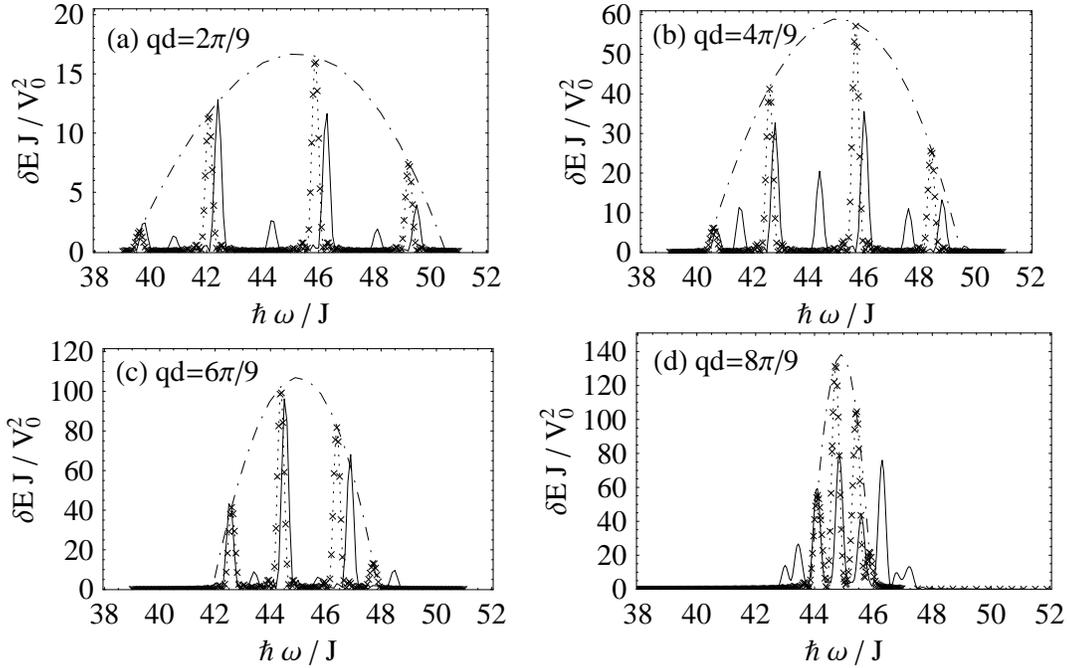}}
\end{center}
\caption{ Energy transfer for a homogeneous system at zero
temperature. Solid line exact solution. crosses perturbative
solution (envelope: dashed line). Other parameters: $M=N=9$, $
J\tau/\hbar=20$, $U/J=45$.
 }\label{101}
\end{figure*}

In Fig. \ref{101} we compare the energy transfer as a function of
the Bragg frequency calculated from Eq. (\ref{contr}) to results
obtained by the exact diagonalization of the BHH for two values of
Bragg momenta $q$. In contrast to the superfluid regime
\cite{Menotti2003a}, where Bragg spectroscopy excites only the
quasiparticle state with quasimomentum $q$, in the Mott regime we
observe $M-1$ peaks pertaining to the two dimensional character of
the 1-ph dispersion relation. The Bragg momentum $q$ fixes one
quantum number $R$ but the other can take $M-1$ different values.
In the analytic solution due to the constraint in Eq.
(\ref{contr}), $\widetilde{q}+r$ even, only $(M-1)/2$ of the
possible $M-1$ peaks are present. The constraint is a consequence
of the extra symmetry introduced in the high filling factor
approximation where similar "effective masses" are assumed.

In the analytic solution, Bragg peaks have an overall envelope of
the form of an inverted parabola centered at the gap energy $U$,
with a maximum height proportional to $\sin^2(qd/2)$ and extended
over an interval with an approximated width of $ 4 J (2g+1)
\cos(qd/2)$. The $\cos(qd/2)$ dependence of the width  and
$\sin^2(qd/2)$ dependence   of the height indicate that as $q$
approaches $\pi/d$ the energy transfer is highly peaked around
$U$. This behavior is observed in Fig. \ref{101}, where the
overall width decreases as $q$ approaches $\pi/d$, while the peak
height increases. In spite of the fact that the analytic solution
uses the high filling factor approximation, the dependence on $q$
of the width and peak height of the envelope is in agreement with
the $g=1$ exact energy response as shown in Fig. \ref{101}.

We found no structure around $\hbar\omega = 2U$ in the exact
numerical results. This is consistent with the fact that the
parameters used in this paper lie within the regime of validity of
first-order perturbation theory.

For the form of energy  spectroscopy we consider here,
 there is no fundamental limit to  Bragg pulsed durations in contrast to  momentum-spectroscopy
\footnote{Momentum spectroscopy requires that the pulse length
does not exceed a quarter of the period of the magnetic trap used
to experimentally confine the atoms\cite{Stenger1999a}}. However,
 practical considerations will likely inhibit the resolution of
the individual excitation peaks in current experiments (see
discussion below). In this case  the envelope shown  in Fig.
\ref{101}  will provide a more realistic depiction of the
experimentally observable spectrum.

\section{Validity conditions}
\label{validity}

The treatment we present here is based on linear response. In this
section we indicate its  strict
 validity conditions in terms of the Bragg strength $V_0$ which is
 the relevant experimental parameter.

After the Bragg perturbation is applied, the many-body state is no
longer in the BHH ground state, $|0\rangle$. The transition
probability to an excited state  $|i\rangle$, $|c_i(t)|^2$,
according to first order perturbation theory is given by
\begin{equation}
|c_i(t)|^2~=~V_0^2 |\langle i |\hat{\rho}^{\dagger}_q|0\rangle|^2
\frac{\sin^2((E_i/\hbar-\omega)t/2)}{\hbar^2(E_i/\hbar-\omega)^2},
\end{equation}
\noindent where  the  eigenenergies $E_i$ of the states are
measured with respect to the ground state energy. The validity of
linear response requires the total excited state population at the
conclusion of the Bragg perturbation to be small compared to
unity:
\begin{equation}
\sum_{i \neq 0}|c_i(T_p)|^2 \ll 1. \label{valid}
\end {equation}

\noindent Deep in the Mott regime the response of the system is
dominated by the $M$ excited states $|\Phi_i^{(1)}\rangle$.
Because all these states have energies $E_i^{(1)}$ approximately
given by $U$, the maximum transfer energy possible is of order
$U$. The validity of linear response constrains the total imparted
energy to be much less than $U$ and the heating rate $\delta
E/\delta t= U\sum_{i \neq 0}|c_i(t)|^2 /t$ due to the Bragg
perturbation to be much less than $U/T_p$.

 It was previously shown that the $M$
excited states $|\Phi_i^{(1)}\rangle$, have an energy spread given
by  $4J(2g+1)\cos(qd/2)$ (Eq.~ (\ref{Spect})), and matrix elements
given by: $|\langle \Phi_i^{(1)} |\hat{\rho}^{\dagger}_q|
\Phi_0^{(1)}\rangle| \propto J \sin(qd/2) \sqrt{32 g(g+1)}/U $,
(Eq.~ (\ref{contr})). The average separation between two
consecutive states is of order $\Delta E \sim 4J
(2g+1)\cos(qd/2)/M$. Individually resolving the different lines
will require one to apply the Bragg pulse for a time of order
$T_p^{(s)} \gtrsim M h /(4J(2g+1)\cos(qd/2)) $. The  validity of
linear  response, Eq.~({\ref{valid}}), therefore requires that
\begin{equation}
V_0\ll \frac{U}{ M }\cot\left(\frac{qd}{2}\right) .
 \end{equation}

\noindent For the parameters of Ref.\cite{Paredes2004}, where
${}^{87}$Rb atoms are trapped in a lattice of depth $18.5 E_R$, the
tunneling time $h/J$ is about $0.1 s$. The number of occupied wells
is $M \sim 20$, with a filling factor $g\sim 1$. Resolving a single
peak would require a Bragg pulse of duration $T_p^{(s)} \sim 0.2 s$.
With these conditions linear response is valid if $V_0 \ll 0.015
\cot(qd/2) E_R$. The acceptable heating rate is much less than $1.8
E_R/s$.

If the duration of the applied perturbation is $T_p \leq
T_p^{(s)}$, excited states will not be individually discernible.
Near resonance ($\hbar \omega\approx U$), for pulse durations
smaller than the inverse  bandwidth: $T_p <T_p^{(e)}$,
$T_p^{(e)}=h/(4J (2g+1) \cos(qd/2))$, all states will be
resonantly excited. If $T_p\sim T_p^{(e)}$  the validity of linear
response requires
 \begin{equation}
V_0 \ll \frac{U}{\sqrt{M}}\cot\left(\frac{qd}{2}\right).
 \end{equation}

\noindent where the factor of $\sqrt{M}$, accounts for the
contribution from all $M$ excited states. For the parameters given
above this inequality implies $V_0~\ll~0.07 \cot(qd/2) E_R$. Here,
the acceptable heating rate is much less than $36 E_R/s$.

\bigskip

 We note that in the superfluid regime the uncorrelated
nature of the system allows for  a less stringent validity
condition to hold: it is only required the amount of excited atoms
to be small compared to the condensate population.

\section{ Finite temperature case}
 \label{fin}

\begin{figure*}[tbh]
\begin{center}
\leavevmode {\includegraphics[width=4.0 in]{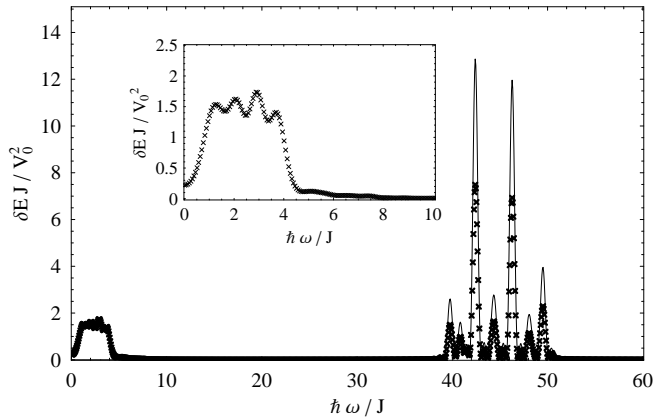}}
\end{center}
\caption{ Energy transfer for a  system at finite temperature. Solid
line $k_BT/U=0.007$, crosses $k_BT/U=0.21$. The other parameters are
the same as those in Fig.\ref{101} }\label{103}
\end{figure*}

It is well known in the literature (see for example
\cite{Brunello2001a}) that in the superfluid regime, Bragg
spectroscopy is not an appropriate tool for probing the temperature
of the system. The reason is that even though $S(q,\omega)$ is
temperature dependent, experimental observables such as the energy
transfer depend  on $\chi(q,\omega)$ which  is very weakly
temperature dependent. This is not the case deep in the Mott
insulator regime. In a translationally invariant lattice all the
1-ph excitations have an energy separation of order $U$ from the
ground state and a splitting between them of order $J$. If the
temperature is  $k_BT \lesssim U/3$, it is still valid to restrict
the Hilbert space to the one spanned by the 1-ph excitations. In
this regime $\chi(q,\omega)$
 (see Eq. (\ref{impene})) can be written as:
\begin{equation}
\chi(q,\omega) =\frac{(1-e^{-\beta U})}{\mathcal{Z}}
\chi(q,\omega)|_{T=0}+\frac{1}{\mathcal{Z}}\chi^{ph}(q,\omega_{ij})
\label{hsq}
\end{equation}
\begin{equation}
\chi^{ph}(q,\omega_{ij})\equiv\sum_{i,j>i}(e^{-\beta
E_i^{(1)}}-e^{-\beta E_j^{(1)}})(f_q(\omega_{ij})-
f_{-q}(-\omega_{ij}))
 \end{equation}

\noindent where the sum runs over the states in the one
particle-hole band, and $\mathcal{Z} \approx 1+M(M+1) e^{-\beta U}$.
The first term in Eq.(\ref{hsq}), proportional to
$\chi_q(\omega)|_{T=0}$, causes a thermal reduction of the zero
temperature response. The second term, which scales like $e^{-\beta
U}\beta J/\mathcal{Z}$ \footnote {Notice that due to the the zeroth
order
 degeneracy  of the $1-ph$ subspace, $\left| \left\langle i\right| \hat{\rho}%
_{q}\left| j\right\rangle \right| ^{2}\sim1.$}, makes  the system
sensitive to low energy Bragg perturbations  at frequencies resonant
with the energy difference between two 1-ph excitations. The factor
$ e^{-\beta U}$ suppresses the observability of these thermal
effects for  $k_BT<U/5$.

In Fig. \ref{103}  we plot the energy transfer as a function of
the Bragg frequency, as calculated from exact diagonalization of
the BHH for two different temperatures. The figure shows that for
temperatures $k_BT/U\gtrsim 1/5$  the height of the zero
temperature peaks around $U$ decreases, while low frequency peaks
appear. The presence of these low-frequency peaks is therefore a
signature of finite temperature. In particular, if peaks around
$U$ are observed in absence of low frequency response, the
temperature is lower than $U/(5 k_B)$. While this analysis does
not provide a precise determination of the temperature, it is
still useful because it shows that Bragg spectroscopy is sensitive
to temperatures of order of the interaction energy.  Current
experimental techniques do not provide any information on the
scale of $U$. In fact, in current experiments temperature
measurements rely on the analysis of atomic interference patterns
after a certain time of flight following the release of atoms, and
the measurement precision is of the order of the energy spacing to
the second lattice band, which is typically one order of magnitude
larger than $U$.
\section{Final Remarks}
\label{con}

 In recent experiments \cite{Stoferle2004a},  Bragg
spectroscopy was performed using a setup where the Bragg momentum
equals the lattice momentum  and response was observed. Our
present analysis, in agreement with previous
ones(\cite{Stoof2004},\cite{Menotti2003a}), predicts no response
for $q=2\pi/d$. Using similar perturbative techniques as the ones
described here, we extended our calculation to inhomogeneous
systems with a strong harmonic magnetic confinement. We also found
no scattering for $q=2\pi/d$ in these systems. The fact that
neither the inhomogeneity nor the finite size of the system are
responsible for the observed signal suggests that nonlinear
effects are the most plausible explanation for the experimental
results.

In summary, we have shown that Bragg spectroscopy can be a suitable
experimental tool for characterizing the Mott insulator phase. By
measuring the transfer energy at different Bragg momenta we proved
it is possible to get information about the excitation spectrum:
Bragg peaks are centered around the characteristic Mott excitation
gap and are contained in an interval whose width  is proportional to
the 1-ph excitation band width. Their average height  is maximized
when the Bragg momentum approaches $\pi/d$. Finally looking at the
low frequency response we showed that Bragg spectroscopy is
sensitive to temperatures of order of the Mott gap.

\section{Acknowledgments}
This work was supported  by an Advanced Research and Development
Activity (ARDA) contract and by the U.S. National Science
Foundation under grant PHY-0100767.

%%%%%%%%%%%%%%%%%%%%%%%%%%%

\end{document}